# Direct observation of micron-scale ordered structure in a two-dimensional electron system

I. J. Maasilta[†], Subhasish Chakraborty, I. Kuljanishvili, and S. H. Tessmer
*Department of Physics and Astronomy, Michigan State University,
East Lansing, MI 48824*

M. R. Melloch
*Department of Electrical Engineering, Purdue University,
West Lafayette, Indiana 47907*

[†]*Current Address: Dept. of Physics, P.O. Box 35, FIN-40014, Univ. of Jyväskylä, Finland*

We have applied a novel scanned probe method to directly resolve the interior structure of a GaAs/AlGaAs two-dimensional electron system in a tunneling geometry. We find that the application of a perpendicular magnetic field can induce surprising density modulations that are not static as a function of the field. Near six and four filled Landau levels, stripe-like structures emerge with a characteristic length scale ~2 μm. Present theories do not account for ordered density modulations on this scale.

Two-dimensional electron systems (2DES) formed in GaAs/AlGaAs heterostructures represent an ideal laboratory to study many-particle physics in lower dimensions [1]. Under various conditions, electron-electron interactions may result in ordered inhomogeneous states. The Wigner crystal represents a classic example, arising from direct Coulomb repulsion at sufficiently low density, temperature, and disorder [2]. In an applied perpendicular magnetic field, more general charge density wave (CDW) ground states are possible. The magnetic field $B$ quantizes the kinetic energy into discrete, degenerate Landau levels; each spin-resolved level holds a density of $eB/h$ electrons. At sufficiently high Landau level filling, Hartree-Fock based theories predict a CDW ground state of length scale on the order of the cyclotron radius, resulting from the competition between Coulomb repulsion and exchange attraction [3]. Recent transport measurements of high-mobility samples show that highly anisotropic dissipation occurs for more than four filled Landau levels, $\nu>4$, with the uppermost level near half filling [4]. Unidirectional CDW stripes are believed to lie at the heart of these observations [5].

In contrast to transport measurements, scanned probe techniques sensitive to electric fields can provide direct images of GaAs/AlGaAs electronic structure [6]. In this paper, we present an extension of Subsurface Charge Accumulation (SCA) imaging [7, 8], adapted to probe the 2DES interior in a novel tunneling geometry. The measurement works as follows: A tunneling barrier separates the 2D layer and a parallel 3D substrate. An *ac* excitation voltage applied between the substrate and a sharp metal tip locally induces charge to tunnel back and forth between the 3D and 2D layers. A Schottky barrier blocks the charge from tunneling directly onto the tip [9]. The measured signal is the resulting *ac* image charge on the tip electrode, which is proportional to the number of electric field lines terminating on it. In this way, the experiment provides a local measurement of the ability of the 2D system to accommodate additional electrons. The method can be considered as a locally resolved version of the pioneering work of Ashoori and coworkers [10], and Eisenstein and coworkers (in a 2D-2D tunneling geometry) [11], which demonstrated that the tunneling signal into a 2DES is sensitive to both gaps in the energy spectrum at integer Landau level fillings, and a Coulomb pseudogap that is pinned to the Fermi level of the 2D system [12]. At particular magnetic fields, we observe micron-scale features, corresponding to ordered electronic density modulations. To the best of our knowledge, these data represent the first tunneling images of the interior of a GaAs/AlGaAs 2DES.

The sample used for these measurements was an $Al_{.3}Ga_{.7}As$ / GaAs (001) wafer grown by molecular beam epitaxy (MBE), shown schematically in Fig. 1(a). The 2DES is located at a distance of $d_1$=60 nm below the exposed surface, and a distance $d_2$=40 nm above a degenerately doped ($10^{18}$ cm$^{-3}$) GaAs substrate (3D metal). The average electron density in the 2DES

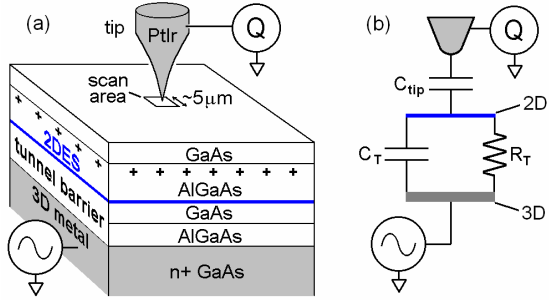

FIG. 1 (color). (a) Schematic of heterostructure sample and measurement technique. The 2DES (indicated in blue) forms in the potential well at the GaAs/AlGaAs interface from electrons provided by silicon dopants (+'s). An AlGaAs tunneling barrier separates the 2DES from a 3D substrate. (b) Equivalent circuit of the sample-tip system.

is $6 \times 10^{11}$ cm$^{-2}$, the low-temperature transport mobility is $\sim 10^5$ cm$^2$/Vs, and the zero magnetic field 3D-2D tunneling rate is approximately 200 kHz. Cryogenic temperatures are achieved by direct immersion in liquid helium-3 at 270 mK.

We position the tip to within a few nanometers of the sample surface using an innovative scanning head that provides a high degree of mechanical and thermal stability [13]. The image charge signal is detected using a circuit constructed from low-input-capacitance high-electron-mobility transistors (noise level 0.01 electrons/√Hz). Most of the signal (~99.8%) corresponds to electric field emanating from areas macroscopically far from the location under the probe [14]. We subtract away this background signal using a bridge circuit. To acquire images, the tip is scanned laterally across the surface without the use of feedback. We apply an excitation voltage of 8 mV rms at a frequency $f$=20 kHz. In addition, we apply a *dc* voltage of 0.6 V to the tip to compensate for the tip-sample contact potential, as measured *in situ* using the Kelvin probe technique [15, 7]; hence the effective *dc* voltage is zero. The spatial resolution of the measurement is roughly 60 nm, limited by the radius of curvature of the chemically etched PtIr tip (50 nm) and by the tip-2DES distance (60 nm) [14]. All images presented here were acquired and are displayed with identical sample orientation.

Fig. 1(b) shows a simple equivalent circuit for the measurement: The local contribution to the signal is modeled as a tip-2DES capacitance $C_{tip}$ in series with the tunneling barrier, represented as a capacitor $C_T$ and resistor $R_T$ in parallel. To account for phase shifts, we define two amplitudes $Q_{in}$ and $Q_{out}$, representing respectively the in-phase and 90° out-of-phase (lagging) components of the image charge per unit excitation voltage. The typical units are attofarads (1 aF=10$^{-18}$ C/V), similar to the earlier SCA work [7].

To establishes the technique's sensitivity to the 2D electron system, Fig. 2(a) shows the measured charging as a function of magnetic field with the tip position fixed (i.e. not scanned), acquired using the same tip as for the subsequently displayed images. Both the in-phase and out-of-phase curves (acquired simultaneously) show clear structure with $1/B$ periodicity, corresponding to integer Landau level fillings. To understand the origin of the in-phase dips and out-of-phase peaks, we consider the charging characteristics of the model equivalent circuit; Fig. 2(b) plots the calculated variation of $Q_{in}$ and $Q_{out}$ with frequency $f$ and resistance $R_T$. The charging components have identical functional dependence on $f$ and $R_T$, with characteristic values of $f_0 = [2\pi R_T(C_T+C_{tip})]^{-1}$ and $R_0 = [2\pi f(C_T+C_{tip})]^{-1}$, respectively. The magnitude of the curves is set by the tip-sample capacitance difference $\Delta C$ between a fully charging and locally non-charging 2DES. For the actual measurements, the characteristic zero magnetic field tunneling rate is $f_0 \approx 200$ kHz, a factor of ten greater than the applied excitation (i.e., $f/f_0 \approx 0.1$). Therefore we focus on the model's low-frequency behavior, $f/f_0 \ll 1$; as shown in the inset, the out-of-phase component is more sensitive to variations in the local tunneling resistance. In contrast, the in-phase component is mostly sensitive to capacitance variations (not shown), including the compressibility contribution to the capacitance [14, 16]. Hence, we conclude that the dips in the measured in-phase curve reflect a reduction in capacitance caused by the diminished compressibility of the 2D system at integer filling. With respect to the out-of-phase curve, the peaks at integer filling likely reflect an increase of the pseudo gap, resulting in an increased tunneling resistance $R_T$ [17]; variations in the in-plane conductivity may also contribute.

Fig. 2(c) presents typical in-phase and out-of-phase charging images far from integer filling. The in-phase data are representative of hundreds of similar images which show charging patterns that are insensitive to magnetic field. We attribute this structure to surface topography, which modulates the geometric capacitance and couples preferentially to $Q_{in}$. This conclusion is supported by comparison to a surface topography measurement (at a different

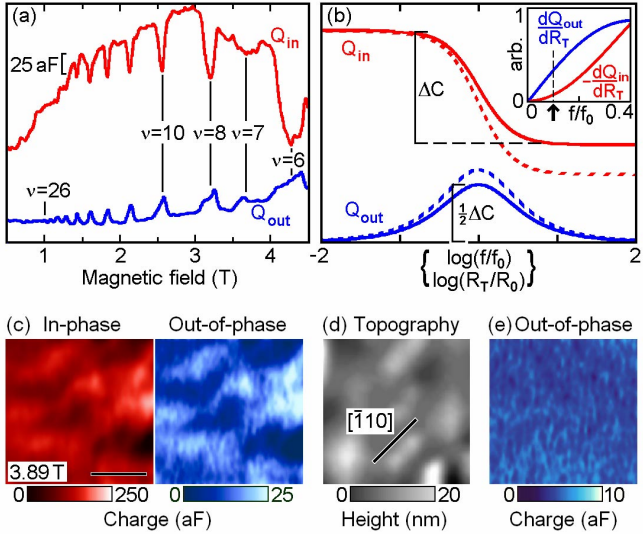

FIG. 2 (color). (a) Fixed-tip measurement. As a function of perpendicular magnetic field, $Q_{in}$ and $Q_{out}$ display clear features at integer Landau level filling $\nu$. (b) Calculated in-phase and out-of-phase charging based on the equivalent circuit. Rigorous modeling for the system considers the distributed nature of the tip-sample capacitance and the effect of charge motion within the 2D plane. The dashed curves show qualitatively enhancements which occur if the in-plane relaxation rates approach the tunneling rate, an effect which may occur near integer filling. (inset) A linear plot of the derivatives of $Q_{in}$ and $Q_{out}$ with respect to tunneling resistance. In the low-frequency limit of this experiment (arrow), the out-of-phase component is more sensitive to $R_T$ variations. (c) Representative scanning images of $Q_{in}$ (left) and $Q_{out}$ (right) with an applied magnetic field far from integer filling of 3.89 T; scale bar length=1 μm. (d) Topographical surface image showing the orientation of the growth features. The image was acquired at room temperature without altering the orientation of the crystal. The elongated mounds point along the [$\bar{1}$10] direction, as indicated by the orientation of the scale bar; scale bar length= 2 μm. (e) Topography-subtracted out-of-phase image corresponding to the same data shown in (c). The images have been filtered to remove nanometer-scale scatter.

location) acquired by operating the microscope in a standard scanning tunneling microscopy mode [13]. As shown in Fig. 2(d), the surface consists of elongated mounds qualitatively similar to the in-phase structure. These surface features point along the [$\bar{1}$10] direction, as indicated, reflecting an MBE growth anisotropy [18].

In contrast to the in-phase images, we find the out-of-phase images can show significant changes with magnetic field. The $Q_{out}$ images also show features arising from surface topography, although this effect is significantly smaller than for the $Q_{out}$ component (as expected in the low-frequency limit). For example, in Fig. 2(c) the out-of-phase color scale is an order of magnitude smaller than the in-phase scale. Because the in-phase component does not change significantly with field, we find that a convenient method to remove the small topographical contribution to $Q_{out}$ is to scale down the corresponding $Q_{in}$ image and then subtract it from $Q_{out}$. Fig. 2(e) shows the same out-of-phase data as Fig. 2(c), where we have subtracted the topography in this way. We see that in this case no significant features remain in $Q_{out}$ discernible above the background noise, as was typical for the out-of-phase images far from integer filling.

As the magnetic field approaches integer filling, while the $Q_{in}$ images remain unchanged, the simultaneously acquired $Q_{out}$ images show completely different behavior – clear and reproducible features appear that are not correlated with topography or experimental parameters such as scan direction. Fig. 3(a) shows a series of topography-subtracted $Q_{out}$ images acquired sequentially near $\nu=6$ at the fields indicated by the small arrows; very similar behavior was observed near $\nu=4$. Similar to the out-of-phase peaks of Fig. 2(a), the imaged $Q_{out}$ structure likely result from local variations of the pseudo gap and/or in-plane conductivity. In either case, we can identify the structure as originating from modulations in the 2DES density ~5% which bring the system slightly closer and further from integer filling as a function of lateral position [14].

Some of the $Q_{out}$ features appear as randomly situated micron scale droplets, such as the bright feature in the upper right corner of the 4.38 T image. These are consistent with droplets and potential contours observed in references [7] and [8], which utilized direct ohmic contacts to the 2DES (i.e. non-tunneling experiments). Following this work, we interpret the droplets as originating from static density modulations induced by the disorder potential. This view is also supported by observations we have made of random features that reappear and evolve in a similar manner at successive integer fillings [18].

In contrast to references [7] and [8], the images exhibit additional structures with orientational order, superimposed with the random droplets. To clearly demarcate these structures, we apply a standard approach based on the 2D Fourier transforms:

$F(k_x, k_y) = \mathcal{F}[Q_{out}(x,y)]$, where $k_x$ and $k_y$ are the wave vector coordinates. The topography has been subtracted from the $Q_{out}$ data as described above. Because we are interested in the spatial patterns and not the shift in the average value (this information is already given by the fixed-tip curves), we have subtracted a constant from each image to set the average value to zero. No additional processing or filtering was performed on the data prior to taking the Fourier transforms.

To further emphasize orientational order, we have also calculated the angular spectrum $A(\theta)$ by integrating the Fourier power spectrum along radial lines oriented at angle $\theta$, $A(\theta) = \int_{k_{min}}^{k_{max}} |F(k\cos\theta, k\sin\theta)|^2 \, dk$, as shown in Fig. 3(b). The integration limits, $k_{min}$ and $k_{max}$, set the range of wave vectors to be included; of course, they can be expressed in terms of cutoff wavelengths: $k_{min}=2\pi/\lambda_{max}$ and $k_{max}=2\pi/\lambda_{min}$. Fig 4(a) presents three sets of unfiltered Fourier transforms and angular spectra corresponding to $Q_{out}$ images of Fig. 3(a). We see that the transform at 4.14 T has developed clear maxima at relatively short wave vectors. To highlight the corresponding range of wavelengths, the angular spectra are calculated choosing cutoffs of $\lambda_{min}=1$ μm and $\lambda_{max}=3$ μm, displayed directly below each transform image. We see that the 4.14 T angular spectrum shows the strongest structure at these short wave vectors; the prominent peak indicated by the red arrow results directly from the maxima in the transform image.

Fig. 4(b) compares the 4.14 T spectrum (near ν=6) to a similar observation obtained at a different location and at a field of 6.75 T (near ν=4). Fig. 4(c) shows the corresponding $Q_{out}$ images, where the arrows point to

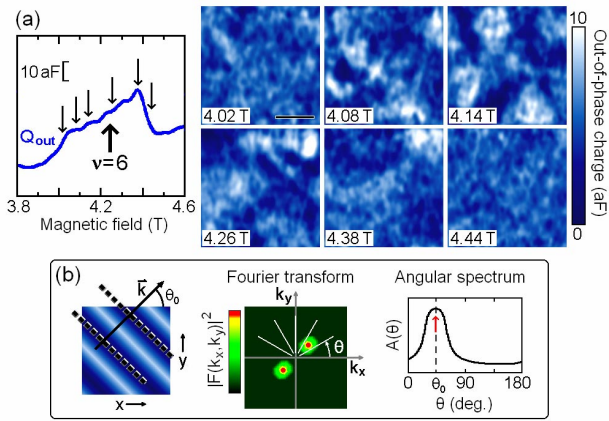

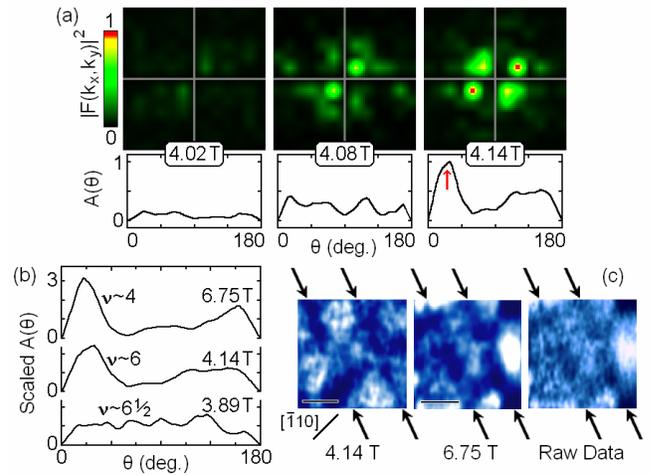

FIG. 3 (color). (a) Topography-subtracted out-of-phase data near six filled Landau levels. The fixed-tip curve to the left shows the ν=6 $Q_{out}$ peak with arrows indicating the fields of the displayed scanning images. The displayed images have been filtered to remove nanometer scale scatter; scale bar length=2 μm. The crystal orientation is identical to Figs. 2(c) and 2(d). (b) Schematic of Fourier analysis procedure to identify orientationally-ordered structure. The parallel lines in the direct image (indicated with dashes) can be described with wave vector $\vec{k}$, giving maxima in the Fourier transform power spectrum at $\pm\vec{k}$ (shown in red). The orientation of the wave vector $\theta_0$ is seen as a peak in the angular spectrum, calculated by integrating the transform image along a series of radial lines at angle θ.

FIG. 4 (color). (a) Fourier power spectra and angular spectra corresponding to three $Q_{out}$ images of Fig. 3(a), normalized to the peak value at 4.14 T. The power spectra are displayed over the range ±6.67 μm$^{-1}$; the magnitude of the wave vector corresponding to the 4.14 T maxima (red) is 2.4 μm$^{-1}$, or a wavelength of 2.6 μm. (b) Angular spectra from three different sample locations and magnetic fields. Similar peaks were observed near four (6.75 T) and six (4.14 T) filled Landau levels. For comparison, a spectrum far from integer filling (3.89 T, corresponding to Fig. 2(e)) is included. In this plot, each curve is scaled to have an average value of 1.0, which serves to enhance the weak 3.89 T curve. The minimum at 0 and 180° is an experimental artifact arising from the scan direction. (c) The topography-subtracted $Q_{out}$ images at 4.14 T and 6.75 T with arrows pointing to the structure that yields the prominent Fourier peaks. Also included is the 6.75 T data for which no filtering was applied. The $[\bar{1}10]$ crystalline direction is indicated; scale bar length=2 μm.

the stripe-like modulations that form the prominent peak in the angular spectra. As an example of the raw data, we also show the 6.75 T out-of-phase image with no filtering. These lines roughly point along the [110] direction. The images show evidence of weaker modulations in other directions. We have acquired dozens of $Q_{out}$ images near ν=6 and ν=4 that show similar micron scale structure. In contrast to the droplet features that mirror static potential variations, the stripe-like structure appears to form spontaneously. Unlike the droplets, these structures shift position and evolve dramatically with magnetic field, as can be seen in Fig. 3. The high sensitivity to magnetic field precludes tunneling barrier nonuniformity as the origin. In addition, we note that scanning tunneling microscopy and atomic force microscopy measurements show no evidence of atomic step edges oriented along [110]. We have not observed such structure near ν=5 nor ν=3. (The 10 T limit of our superconducting solenoid prevents probing Landau level fillings less than ν=3 for this sample.) We have not found evidence of ordered structure on the predicted cyclotron radius length scale, ~100 nm.

In summary, we have obtained the first direct images of ordered structure within the interior of a GaAs/AlGaAs 2DES using a novel scanned probe method. At magnetic fields near four and six filled Landau levels, we observe density modulations that exhibit clear orientational order, roughly along the [110] crystalline direction. The features are suggestive of micron-scale stripe-like states. Because the spacing between features is a substantial fraction of the experimental scan range, we can only give a rough value for the characteristic length scale and cannot speak to the periodicity of the structure. It is tempting to equate the observations with states similar to theoretical high-Landau-level charge density wave states [3]. However, the length scales represent a major discrepancy: the theories predict a characteristic length set by the cyclotron radius ~100 nm, much smaller than the observed structures ~2 μm. It is possible that the physical origin of the modulations is similar to the CDW theories, but that disorder causes a larger length scale to be selected. Alternatively, the structures may arise from an entirely different mechanism. Clearly, more theoretical work is needed.

We thank R. C. Ashoori, J. Bass, S.J.L. Billinge, H.-B. Chan, V. J. Goldman and S. Urazhdin for discussions and assistance. We acknowledge enlightening conversations with M. I. Dykman, A. H. MacDonald, and J. Sinova with regard to the interpretation of these data. This work was supported by the National Science Foundation grant DMR00-75230. SHT acknowledges support of the Alfred P. Sloan Foundation.

---